\documentclass{aa}
\usepackage[varg]{txfonts}
\usepackage{natbib}
\usepackage{url}
\usepackage{hyperref}
\usepackage{xcolor}
\usepackage{ulem}
\usepackage{orcidlink}
\usepackage{xspace}

\usepackage{upgreek}
\usepackage{graphicx}   
\usepackage{multirow}
\usepackage[switch]{lineno}
\usepackage{subcaption} 

\newcommand{\src}{{4U~1954$+$319}\xspace}
\newcommand{\nustar}{\textit{NuSTAR}\xspace}
\newcommand{\ixpe}{\textit{IXPE}\xspace}

\newcommand{\gaia}{\textit{Gaia}\xspace}

\def\flux{erg\,s$^{-1}$\,cm$^{-2}$\xspace}
\def\lum{erg\,s$^{-1}$\xspace}

\begin{document}

\title{First detection of X-ray polarization from \\ the long-period X-ray pulsar 4U~1954$+$319}

\titlerunning{X-ray polarization of 4U~1954$+$319}

\authorrunning{Salganik, A., et al.}

\author{Alexander~Salganik\inst{\ref{in:UTU}}\orcidlink{0000-0003-2609-8838}, 
Lingda~Kong\inst{\ref{in:TUB}}\orcidlink{0000-0003-3188-9079}, 
Sofia V. Forsblom\inst{\ref{in:UTU}}\orcidlink{0000-0001-9167-2790}, 
Menglei Zhou\inst{\ref{in:TUB}}\orcidlink{0000-0001-8250-3338}, 
Honghui Liu\inst{\ref{in:TUB}}\orcidlink{0000-0003-2845-1009}, \\ 
QingChang Zhao\inst{\ref{in:IHEP}, \ref{in:UCAS}}\orcidlink{0000-0001-9893-8248},
Sergey~S.~Tsygankov\inst{\ref{in:UTU},\ref{in:TUB},\ref{in:IHEP}}\orcidlink{0000-0002-9679-0793}, 
Andrea Santangelo\inst{\ref{in:TUB}}\orcidlink{0000-0003-4187-9560}, 
Juri Poutanen\inst{\ref{in:UTU}}\orcidlink{0000-0002-0983-0049}}
\institute{
Department of Physics and Astronomy, 20014 University of Turku,  Finland 
\label{in:UTU} \\ \email{alsalganik@gmail.com}
\and
Institut für Astronomie und Astrophysik, Universität Tübingen, Sand 1, 72076 Tübingen, Germany
\label{in:TUB}
\and
Key Laboratory of Particle Astrophysics, Institute of High Energy Physics, Chinese Academy of Sciences, Beijing 100049, China \label{in:IHEP}   
\and 
University of Chinese Academy of Sciences, Chinese Academy of Sciences, Beijing 100049, China
\label{in:UCAS}
}

\date{Received 5 December 2025 / Accepted 19 March 2026}

\abstract{We report the first detection of X-ray polarization with the \textit{Imaging X-ray Polarimetry Explorer} from the X-ray pulsar (XRP) \src.  
The source belongs to an extremely rare class of systems in which a slowly rotating neutron star accretes from the dense wind of a red supergiant companion. 
We detect coherent pulsations at $P_{\rm spin}=5.49\pm0.05$~h, which is one of the longest spin periods known among XRPs.
While the phase-averaged analysis shows no significant polarization, with a 99\% confidence minimum detectable polarization (MDP$_{99}$) of 4.9\% in the 2--8 keV band, the phase-resolved analysis shows  a single interval at pulse maximum in which the polarization degree (PD) exceeds the MDP$_{99}$, yielding ${\rm PD}=10.2_{-3.0}^{+3.1}\%$. 
The polarization angle (PA) exhibits a smooth $\approx150\degr$ rotation over the pulse, and a joint evaluation of all phase bins yields an overall detection significance of $3.3\sigma$.
Using the rotating vector model, we identify a geometric solution that reproduces the observed PA variation. From this model, we infer a phase-independent ${\rm PD}$ of $6.1\pm1.1$\% in the 2--8~keV band.}


\maketitle

\section{Introduction}

Neutron stars (NSs) are among the most extreme objects in the Universe, with surface magnetic fields spanning $\sim10^{8}$ G in recycled millisecond pulsars, $\sim10^{11}$--$10^{13}$ G in ordinary pulsars, and up to $\sim10^{14}$--$10^{15}$~G in magnetars. 
In accreting binary systems, they manifest as X-ray pulsars (XRPs), which serve as natural laboratories for studying accretion physics, magnetospheric interactions, and quantum electrodynamic effects in strong fields \citep{MushtukovTsygankov2024}. Their X-ray emission is fueled by the accretion of matter from the companion star, producing hot spots or accretion columns near the NS magnetic poles. 
The structure and geometry of this emission region are primarily governed by the magnetic field strength and the accretion rate. XRPs occur in both low-mass X-ray binaries \citep[LMXBs;][]{Avakyan2023} and  high-mass X-ray binaries \citep[HMXBs;][]{Neumann2023}.

Among X-ray binaries, \src\ represents a particularly unusual case. 
The donor star was originally classified as an M4--5\,III giant \citep{Masetti2006}, which led to its identification as a symbiotic X-ray binary (SyXRB), a subclass of LMXBs where an NS accretes from the dense wind of an evolved late-type giant \citep[see][for a review]{Bahramian2024}. 
However, later observations revealed that the donor is a red supergiant  (RSG) of spectral type M4\,I (\citealt{Hinkle2020}), implying that \src\ is an HMXB rather than a SyXRB. 
While most known HMXBs host blue supergiant or Be-type donors, \src\ stands out as one of only two confirmed Galactic systems containing an NS accreting from an RSG, the other being SWIFT~J0850.8$-$4219 \citep{De2024}. 
The updated classification and revised stellar parameters indicate a donor mass of $\sim$$9\,M_{\odot}$, a radius of $\sim$$590\,R_{\odot}$ \citep{Hinkle2020}, and a \gaia~DR3 distance of $d = 3.89^{+0.33}_{-0.28}$~kpc \citep[][their Table~2]{WangLi2025}. 
These properties imply a binary separation of at least 5~au and an orbital period longer than three years \citep{Hinkle2020}.

 \citet{Corbet2006, Corbet2008} discovered pulsations from \src with a period of about 5~h, establishing it as one of the slowest known XRPs.
The NS spin period exhibits substantial long-term variations of more than 10\% with alternating spin-up and spin-down episodes \citep{Marcu2011, Enoto2014}. 
The pulse profile in the 2--8~keV range of the \textit{Imaging X-ray Polarimetry Explorer} (\ixpe) shows a complex morphology characterized by two broader maxima and multiple subpeaks with a pulsed fraction (PF) of approximately 60--80\% \citep{Enoto2014}. 
 \citet{Bozzo2022} used hardness-ratio resolved spectroscopy with \textit{X-ray Multi-Mirror Mission (XMM-Newton)} and \textit{Nuclear Spectroscopic
Telescope Array (NuSTAR)} observatories to reveal variability consistent with accretion from a clumpy RSG wind.

The nature of \src\ makes it not only an evolutionary outlier among X-ray binaries, but also a key test case for accretion theory. 
The low bolometric luminosity (up to $\sim10^{36}$~\lum; \citealt{Marcu2011, Enoto2014, Bozzo2022}), the extremely long spin period, and the wind-fed accretion scenario are difficult to reconcile with standard angular momentum transfer models such as the magnetically threaded accretion disk model of  \citet[see, e.g., \citealt{2024MNRAS.533..386M}]{GhoshLamb1979}.  
In contrast, the subsonic settling accretion model \citep{Shakura2012}, especially when modified to include stochastic reversals of the accreted angular momentum \citep{2024MNRAS.533..386M}, explains the observed properties without requiring previously suggested magnetar-strength fields \citep{Bozzo2022}.

Although spectral and timing analyses have constrained the accretion regime of \src, they cannot reveal the orientation of the magnetic field or the overall emission geometry.
X-ray polarimetry with IXPE now makes this possible and provides direct insights into the system’s accretion configuration.

In this work, we report the first detection of X-ray polarization in an XRP with an RSG companion, obtained for \src with \ixpe. 
We detected the polarization signal in the phase-resolved analysis, while the phase-averaged emission showed no significant polarization. 
 \citet{Atel17464} presents the preliminary results.
The remainder of this paper is organized as follows.  
Section~\ref{sec:observations} provides an overview of the observational methods and data sets used.
The main results are presented in Sect.~\ref{sec:results}.
Finally, Sect.~\ref{sec:rvm} presents the constraints on the pulsar geometry, and Sect.~\ref{sec:conclusions} summarizes our findings.

\section{Observations and data reduction}
\label{sec:observations}
\subsection{\ixpe}
The \ixpe observatory (\citealt{Weisskopf2022}) carries three identical detector units (DUs), each equipped with a gas pixel detector \citep{Baldini2021,Soffitta2021}.  
Observations with the \ixpe covered \src from 16 to 21 October 2025 (ObsID 04251201), for a total exposure of about 194~ks per DU.
During this observation, data from DU2 were unavailable due to pixel failures that occurred in April 2025, so the analysis relied solely on photons detected by DU1 and DU3.

We extracted spectral and polarimetric products using the \texttt{ixpestartx}\footnote{\url{https://heasarc.gsfc.nasa.gov/docs/ixpe/analysis/contributed/ixpestartx.html}} task, which uses the standard HEASoft v6.36 together with the most recent \ixpe calibration files (CALDB v.20250225). 
We adopted a circular extraction region of 80\arcsec\ radius centered on the source (R.A. = $19^\mathrm{h}55^\mathrm{m}42^\mathrm{s}$, Dec. = $+32\degr05\arcmin49\arcsec$). 
We tested weighted analysis and background subtraction using different annular regions centered on the source and found no significant improvement in the polarization constraints. We therefore adopted an unweighted analysis without background subtraction.
In all event files, we rejected events with a nonzero status column, thereby removing events flagged as instrumental background or otherwise problematic.

We obtained the polarization degree (PD) and polarization angle (PA) parameters using two complementary approaches: (1) a model-independent analysis with the \texttt{ixpepolarization} tool in HEASoft \citep[see also][]{Kislat2015} and (2) spectro-polarimetric fitting with \texttt{xspec}.  
In addition to the phase-averaged analysis, we performed a phase-resolved study. 
We filtered barycenter-corrected (via \texttt{barycorr}) event lists in phase by assigning each photon to the appropriate phase bin. We did not apply a binary motion correction because the complete set of source orbital parameters remains unknown.
For each phase bin, we generated spectra and responses with \texttt{ixpestartx} and then grouped them with \texttt{ftgrouppha}, requiring a minimum of 20 counts per bin and using the same energy bins for $I$, $Q$, and $U$. 
We fit the resulting spectra with \textsc{xspec} v. 12.15.1  \citep{xspec1996}.

\subsection{\nustar}
\nustar comprises two identical, co-aligned X-ray focal plane modules, FPMA and FPMB \citep{Harrison2013}, which provide sensitive coverage in the 3--79~keV energy band. Observations were performed on 2025 October 16 (ObsID 91102340002) and on 2025 October 18 (ObsID 91102340004), hereafter referred to as NuObs1 and NuObs2, respectively. We extracted the source events using a circular region with a 50\arcsec\ radius centered on the source position. We chose a 120\arcsec\ radius background region to optimize the signal-to-noise ratio, particularly at higher energies. 

We reduced the \nustar observations following the standard processing guidelines.\footnote{\url{https://heasarc.gsfc.nasa.gov/docs/nustar/analysis/nustar_swguide.pdf}}  The data were processed using \textsc{HEASoft} v6.36 and CALDB version 20251006 (including the clock correction file 20100101v215). We produced cleaned event files with the task \texttt{nupipeline}. To mitigate background enhancements associated with passages through the South Atlantic Anomaly (SAA), we applied the filtering configuration recommended by the official SAA reports for these sequences, namely \texttt{saacalc=2}, \texttt{saamode=OPTIMIZED} and \texttt{tentacle=no}, which preserves essentially all useful exposure (21.3 and 19.4~ks for NuObs1 and NuObs2, respectively). We extracted the spectra with the \texttt{nuproducts} tool as part of the \texttt{nustardas} pipeline and
grouped them to have at least 20 counts per energy channel.  
Significant stray-light contamination affected FPMB during NuObs1 and NuObs2; therefore, we excluded this module from the spectral analysis.

Throughout this paper we calculate all luminosities assuming the \gaia~DR3 distance of $d = 3.89$~kpc \citep{WangLi2025}. We adopt the abundances of \citet{Wilms2000} for the modeling of photoabsorption.
We perform spectral fits using $\chi^2$ statistics and quote uncertainties at the 68.3\% confidence level (CL; $1\sigma$), unless otherwise stated.

\section{Results}
\label{sec:results}

\subsection{Timing analysis}
\label{sec:timing}
\begin{figure}
    \centering
    \includegraphics[width=\linewidth]{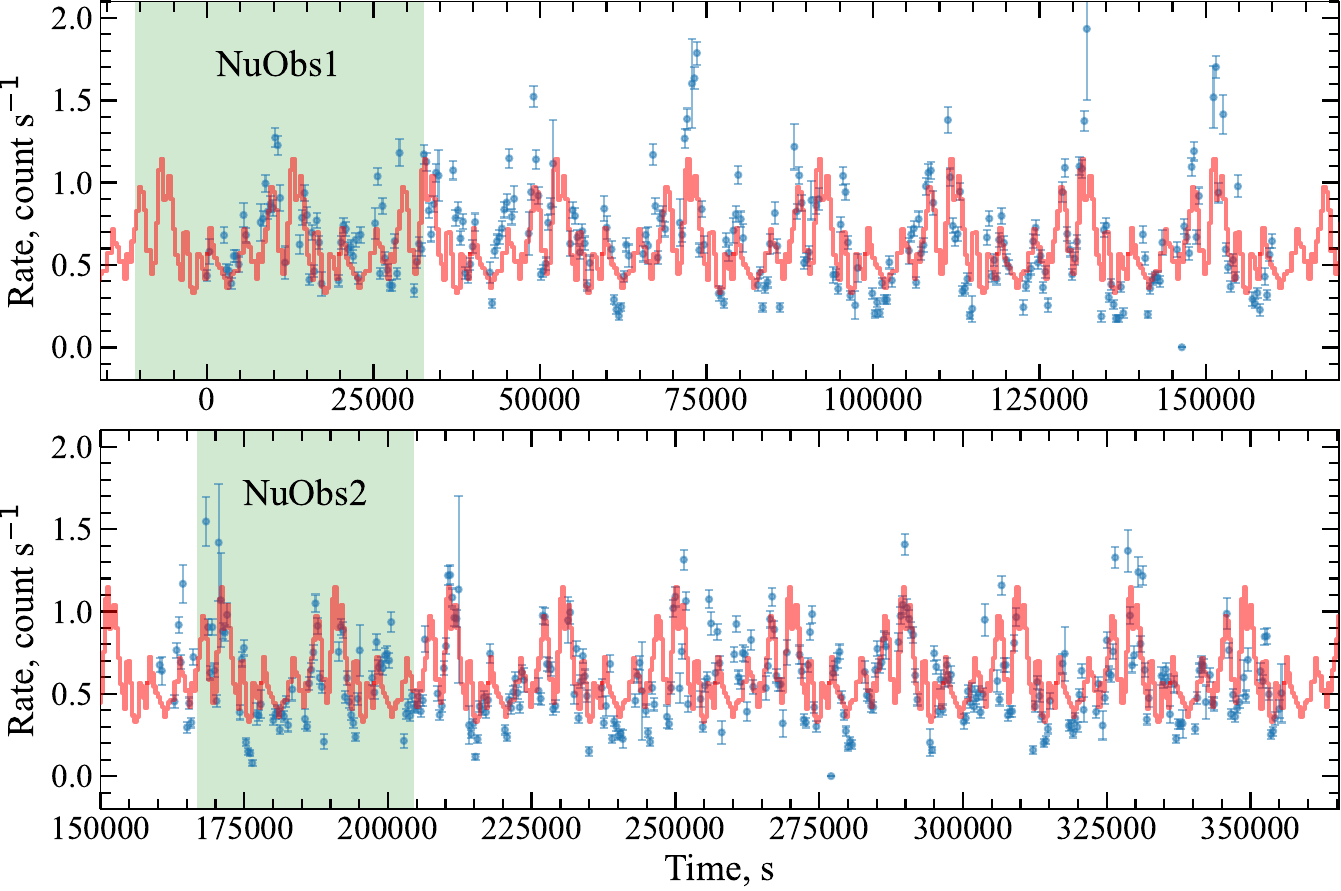}
    \caption{Light curve of \src\ in the 2--8~keV band during the \ixpe\ observation, rebinned to 366~s. The zero time corresponds to the start of the \ixpe\ observation. Shaded regions mark the time intervals of the two \nustar\ observations (NuObs1 and NuObs2). The solid curve shows the pulse profile repeated over the spin period.}

    \label{fig:lc}
\end{figure}

For the timing analysis, we extracted 2--8~keV light curves from the event files of DU1 and DU3. 
Using \textsc{xselect}, we applied an energy filter and selected source photons from the same circular region used for the spectral analysis. 
We produced light curves with a time bin of 1~s, and summed the resulting curves from DU1 and DU3 using the \texttt{lcmath} procedure.  Figure~\ref{fig:lc} shows the resulting summed 2--8~keV light curve.

\begin{figure}
\centering
\includegraphics[width=\columnwidth]{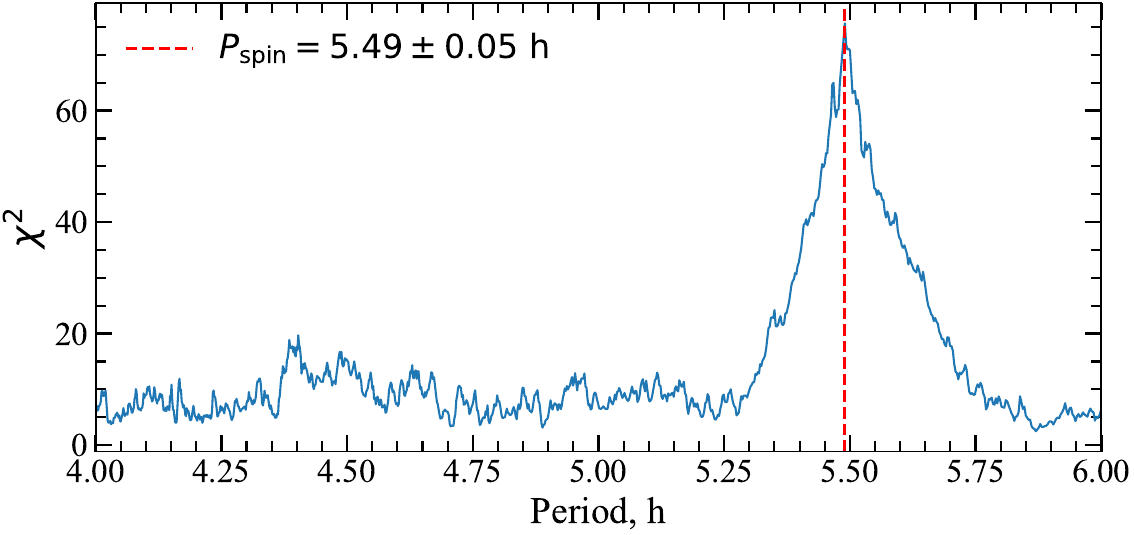}
\caption{$\chi^2$ periodogram of the 2--8~keV \ixpe light curve. 
The best-fit period is indicated by the vertical dashed line.}
\label{fig:period}
\end{figure}

The $\chi^2$ periodogram gives a peak at $5.49\pm0.05$~h (Fig.~\ref{fig:period}).  
We estimated the uncertainty using a bootstrap resampling procedure, randomizing the light curve repeatedly within its statistical uncertainties and fitting the main peak of the periodogram with a Gaussian. 
This period corresponds to the NS spin period of $\sim 5$~h \citep{Corbet2006, Corbet2008} and serves as the basis for the phase-resolved  polarimetric analysis presented below.

\begin{figure}
\centering
\includegraphics[width=0.92\columnwidth]{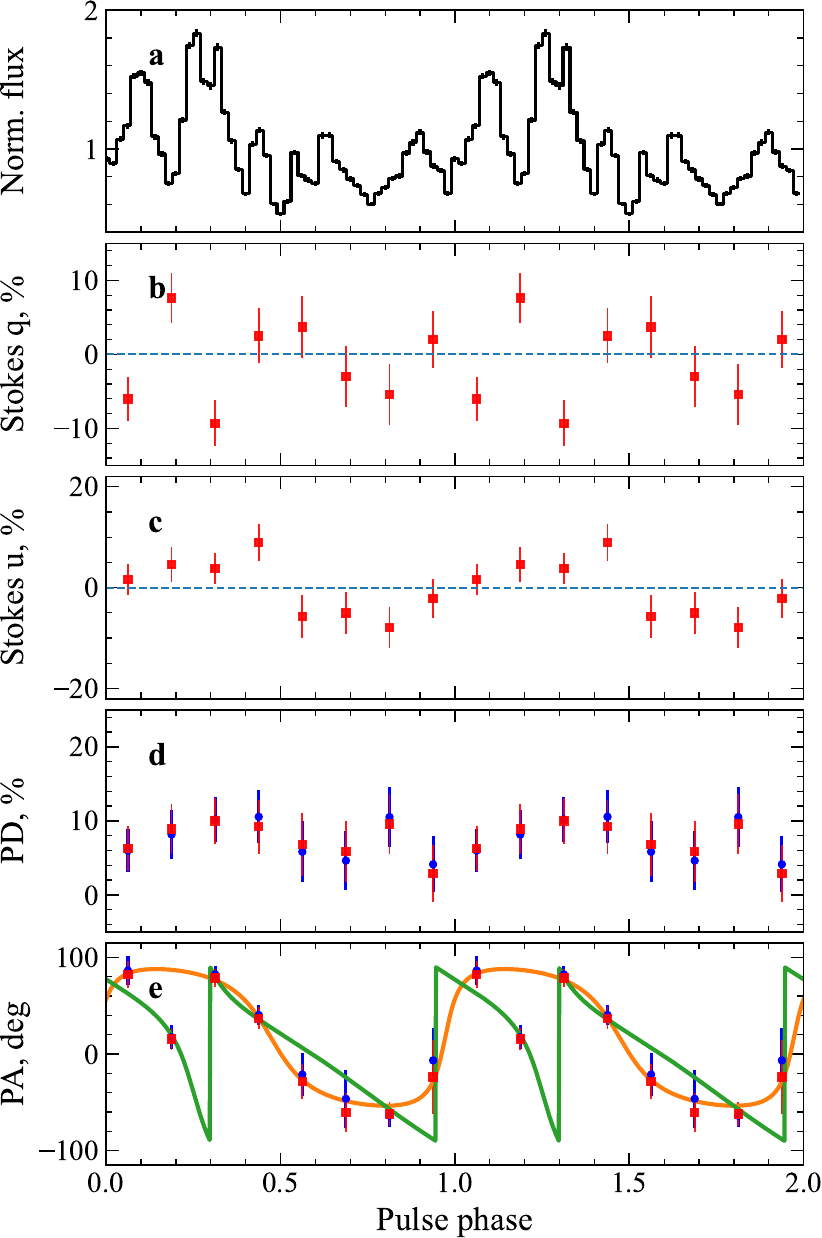}
\caption{Phase-resolved properties of \src over two pulse cycles. 
The PD and PA measurements derive from two methods: blue points show spectro-polarimetric analysis results, and square red symbols show model-independent measurements obtained with \texttt{ixpepolarization}. 
Panels display (a) the normalized pulse profile in the 2--8~keV band derived from the IXPE data, (b)--(c) the Stokes parameters $q$ and $u$, (d) the PD, and (e) the PA.
The PA panel shows two RVM solutions from Table~\ref{tab:rvm_solutions}, corresponding to two distinct posterior clusters in Fig.~\ref{fig:corner_rvm}: Solution 1 (orange) and Solution 2 (green).  }
\label{fig:phase_stack_rvm}
\end{figure}

Folding the 2--8~keV light curve on this period yields a complex pulse profile (Fig.~\ref{fig:phase_stack_rvm}a).  
The profile shows two maxima and multiple subpeaks.  The light curve does not show statistically significant variability that would indicate changes in the spectral state during the \ixpe\ observation. The pulse profile, overplotted on the light curve, follows the data without indicating long-term deviations indicative of additional variability (Fig.~\ref{fig:lc}). This suggests that no strong changes in the dominant polarization properties occur over the course of the observation.

\subsection{Phase-averaged polarimetric analysis}
\label{sec:avg}

We performed two separate phase-averaged joint spectral and polarimetric fits in \textsc{xspec}. Each fit combined the full \ixpe\ dataset with a different quasi-simultaneous \nustar\ observation, namely NuObs1 and NuObs2. In both cases, the \ixpe\ data consisted of six spectra corresponding to Stokes $I$, $Q$, and $U$ from DU1 and DU3. We initially adopted the model
\texttt{const$\times$tbabs$\times$[cutoffpl$\times$polconst+gaussian]},
where \texttt{const} accounts for cross-calibration offsets between instruments (with DU1 fixed to unity), \texttt{tbabs} models interstellar absorption, \texttt{polconst} assumes energy-independent polarization across the \ixpe\ band, and the Gaussian component represents the Fe K$\alpha$ emission around 6.4~keV.

When fitting NuObs1+\ixpe, the baseline model does not fully describe the data ($\chi^2/\mathrm{d.o.f.}=1.09$), leaving systematic residuals in the form of a broad hump at 20--30~keV. To account for this feature, we included an additional \texttt{bbodyrad} component, resulting in the final model
\texttt{const$\times$tbabs$\times$[(cutoffpl+\texttt{bbodyrad})$\times$polconst + gaussian]}.
The additional component has a temperature of $kT \approx 1.5$~keV and an emitting radius of $\sim600$~m, improving the fit to $\chi^2/\mathrm{d.o.f.}=1.06$ and substantially reducing the residual structure. The same extended model yields an acceptable fit for NuObs2+\ixpe\ ($\chi^2/\mathrm{d.o.f.}=0.92$), with no notable residuals in either dataset, including the energy range around the iron line.

\begin{figure}[t]
    \centering
    \includegraphics[width=\linewidth]{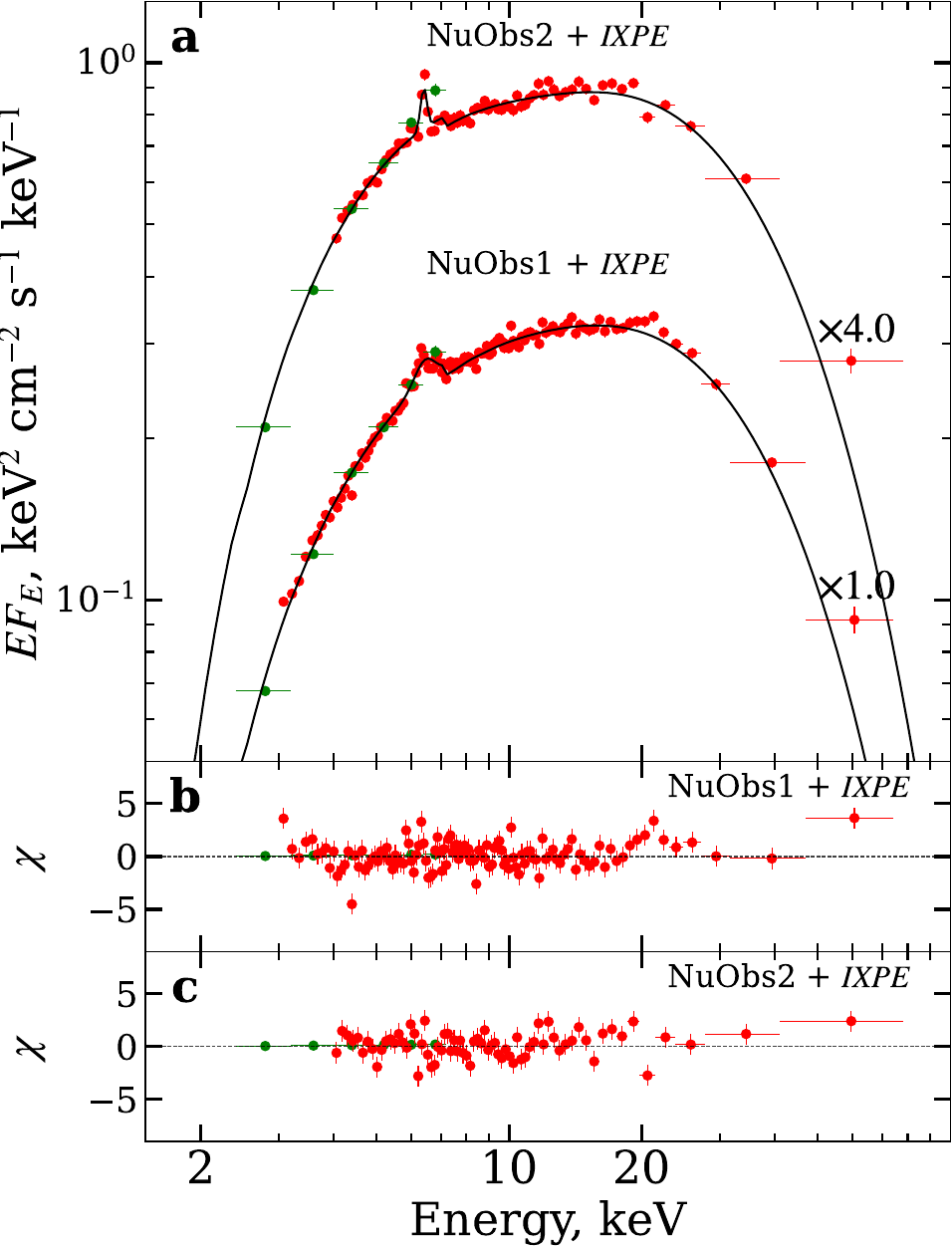}
\caption{
Unfolded energy spectra from the joint \ixpe\ and \nustar analysis.
The \ixpe\ spectra correspond to the Stokes $I$ datasets from DU1 and DU3. Green points denote the \ixpe\ measurements, while red points represent the \nustar\ spectrum. Cross markers indicate the multiplicative scaling factors applied to the spectra for visual clarity.
Panels~(b) and (c) show residuals of the joint fits for NuObs1+\ixpe\ and NuObs2+\ixpe, respectively. 
}
    \label{fig:joint_ixpe_nustar}
\end{figure}

The joint NuObs2+\ixpe\ fit also shows better overall consistency between the two observatories than the NuObs1+\ixpe\ case. We therefore adopted the NuObs2-based parameters as our final phase-averaged spectral--polarimetric results. Our analysis does not significantly detect phase-averaged polarization and constrains it with an upper limit of $5.1\%$ (corresponding to $\mathrm{MDP}_{99}$). Table~\ref{tab:specpol_fit} (see also Fig.~\ref{fig:joint_ixpe_nustar}) lists the best-fitting spectral and polarimetric parameters.

\begin{table}[t]
\centering
\caption{Best-fitting joint phase-averaged spectral and polarization parameters.}
\label{tab:specpol_fit}
\begin{tabular}{lcc}
\hline
\hline
Parameter & \multicolumn{2}{c}{Value} \\
\hline
&  NuObs1 + \ixpe & NuObs2 + \ixpe\\
$N_{\rm H}$ ($10^{22}\,\mathrm{cm^{-2}}$) & $4.3\pm0.4$ & $5.2\pm0.4$ \\
$\Gamma$ & $0.7\pm0.1$ &  $0.9\pm0.1$\\
$E_{\rm fold}$ (keV) & $12.9\pm0.6$ & $14.1^{+0.9}_{-0.8}$ \\
$T_{\rm bbodyrad}$, keV & $1.54\pm0.03$ & $1.41\pm0.03$\\
$R_{\rm bbodyrad}$, m & $590\pm50$& $620\pm50$\\
$E_{\rm Fe\,K\alpha}$ (keV) & 6.4 (frozen) & 6.4 (frozen) \\
$\sigma_{\rm Fe\,K\alpha}$ (keV) & $0.30\pm0.08$ & 0.1 (frozen) \\

$\mathrm{EW}_{\rm Fe\,K\alpha}$ (keV) & $0.09\pm0.01$ & $0.05\pm0.01$ \\

Flux$_{2-8}$ & $4.1\pm0.1$ & $3.3\pm0.1$\\
Luminosity$_{2-8}$  & $7.4\pm0.1$  &$5.9\pm0.1$\\
Flux$_{3-79}$ & $11.8\pm0.1$ & $8.4\pm0.1$ \\
Luminosity$_{3-79}$ &  $21.5^{+0.2}_{-0.1}$ &$15.4\pm0.1$ \\

$\mathrm{PD}$ (\%) & $<5.4$ ($\mathrm{MDP}_{99}$) &  $<5.1$ ($\mathrm{MDP}_{99}$) \\
$\mathrm{PA}$ (deg) & $88\pm19$ & $84\pm22$ \\
$\chi^2/\mathrm{d.o.f.}$ & $796/748$ & $586/639$\\
\hline
\end{tabular}
\tablefoot{Fluxes are given in units of $10^{-10}$~\flux, luminosities in units of $10^{35}$~\lum.}

\end{table}

\begin{table*} 
\centering
\caption{Phase-resolved polarization and spectral parameters of \src\ in the 2--8~keV band, based on model-independent \texttt{ixpepolarization} and spectro-polarimetric \textsc{xspec} analyses.}
\label{tab:phase_ixpepol}
\begin{tabular}{ccccccccc}
\hline\hline
Bin &  Phase range & $N_{\rm H}$ & $\Gamma$ & ${\rm PD}_{\text{ixpepol}}$ & 
${\rm PA}_{\text{ixpepol}}$ & 
${\rm PD}_{\text{xspec}}$ & 
${\rm PA}_{\text{xspec}}$ & 
Significance$_{\text{xspec}}$\\
 &  &($10^{22}~{\rm cm}^{-2}$)&& (\%) & (deg) & (\%) & (deg) & ($\sigma$) \\
 \hline
00 & 0.000--0.125 &$7.5\pm0.7$&$1.2\pm0.1$& $6.2\pm3.0$ & $82\pm14$ & $6.0_{-2.9}^{+3.0}$ & $87\pm15$ &  $2.0$\\
01 & 0.125--0.250 &$4.8\pm0.7$&$0.8\pm0.1$& $8.9\pm3.4$ & $16\pm11$ & $8.1\pm3.4$ & $18\pm12$ & $2.4$\\
02 & 0.250--0.375 &$8.0\pm0.7$&$1.5\pm0.1$& $10.0\pm3.1$ & $79\pm9$ & $10.2_{-3.0}^{+3.1}$ & $83\pm9$ & $3.3$\\
03 & 0.375--0.500 &$6.1\pm0.8$&$1.1\pm0.1$& $9.2\pm3.7$ & $37\pm11$ & $10.6\pm3.6$ & $40\pm10$  & $2.9$\\
04 & 0.500--0.625 &$6.1\pm0.9$&$1.3\pm0.2$& $6.8\pm4.2$ & $-28\pm18$ & $5.8\pm4.1$ & $-21\pm22$ & $1.4$\\
05 & 0.625--0.750 &$6.4\pm0.9$&$1.8\pm0.2$& $5.8\pm4.1$ & $-60\pm20$ & $4.6\pm4.0$ & $-46\pm30$ & $1.1$\\
06 & 0.750--0.875 &$6.8\pm0.9$&$1.8\pm0.2$& $9.6\pm4.1$ & $-62\pm12$ & $10.5\pm4.0$ & $-64\pm11$ & $2.6$\\
07 & 0.875--1.000 &$5.8\pm0.8$&$1.1\pm0.1$& $2.9\pm3.9$ & $-24\pm38$ & $4.1\pm3.8$ & $-7_{-27}^{+33}$ &  $1.1$\\

\hline
\end{tabular} 
\end{table*}

We further investigated the energy dependence of the polarization in \src by dividing the 2--8 keV band into three sub-bands: 2--4, 4--6, and 6--8 keV. 
We performed energy-resolved spectro-polarimetric fits for each energy interval using the same model configuration, with the Stokes $Q$ and $U$ spectra restricted to the corresponding energy bins. 
We could not significantly constrain the PD in any of these bands and therefore report $\mathrm{MDP}_{99}$ upper limits of 5.2\%, 5.2\%, and 6.1\% for the 2--4, 4--6, and 6--8~keV bands, respectively.

\begin{figure}
\centering
\includegraphics[width=\columnwidth]{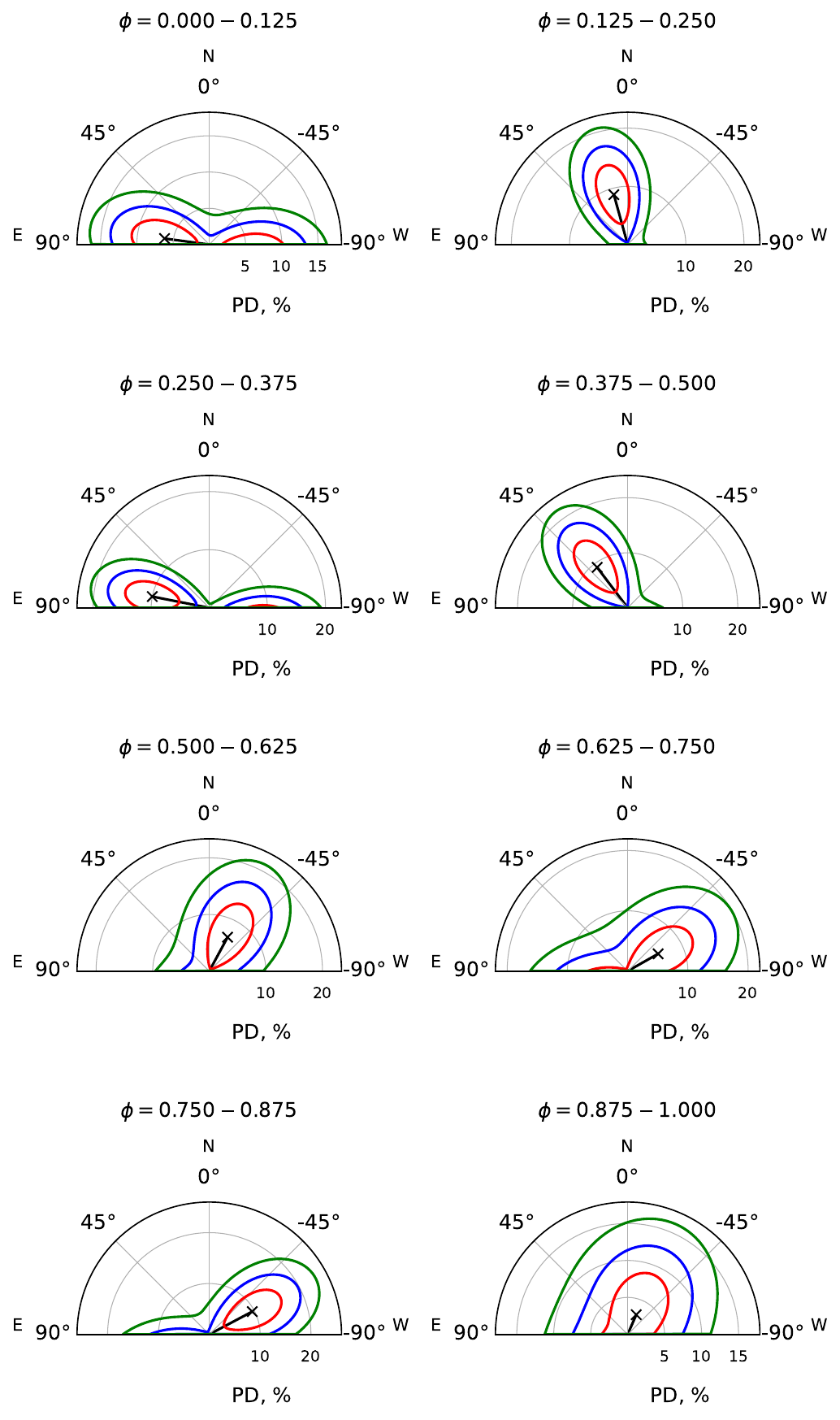}
\caption{Protractor plots showing phase-resolved PD (radius) and PA (azimuth) in the 2--8~keV range. 
Contours indicate 1, 2, and 3$\sigma$ CLs for two parameters, shown in red, blue, and green, respectively.
The cross marks the best-fit point, and the line indicates the polarization direction. }
\label{fig:ixpe_protractor}
\end{figure}

\subsection{Phase-resolved polarimetric analysis}
\label{sec:phres}

We performed a pulse phase-resolved polarimetric analysis by dividing the \ixpe 2--8~keV data into eight equal phase bins using the ephemeris $T_{0}=\mathrm{MJD(TDB)}~60963.927$ and $P_{\rm spin}=5.49\pm0.05$~h (derived from the \ixpe light curve; see Sect.~\ref{sec:timing}). For the spectral-polarimetric analysis in each phase bin, we adopted the \textsc{xspec} model \texttt{const$\times$tbabs$\times$(po$\times$polconst)}, due to the narrow energy range of \ixpe.
For each phase interval, we calculated PA and PD using \textsc{xspec} and \texttt{ixpepolarization} (see Sect.~\ref{sec:avg}).
Table~\ref{tab:phase_ixpepol} and Fig.~\ref{fig:phase_stack_rvm} summarize the phase-resolved polarization results in the 2--8~keV band.
We derived the data points for the normalized Stokes parameters $q$ and $u$ (Figs.~\ref{fig:phase_stack_rvm}b and \ref{fig:phase_stack_rvm}c) solely from the model-independent \texttt{ixpepolarization} method.

The spectro-polarimetric analysis shows that the phase interval $\phi=0.250$--$0.375$ (bin~02) shows a PD exceeding its MDP$_{99}$ threshold. 
In this interval, the PD reaches $10.2_{-3.0}^{+3.1}$\% at $\mathrm{PA}=83\degr\pm9\degr$ (3.3$\sigma$), which corresponds to the pulse-maximum. 
The phase dependence of the PA suggests a smooth rotation across the pulse by $\approx150\degr$, from a minimum of about $-64\degr$ (bin~06) to a maximum of about $87\degr$ (bin~00; see also Fig.~\ref{fig:ixpe_protractor})

Following the \ixpe statistical recommendations,\footnote{\url{https://heasarc.gsfc.nasa.gov/docs/ixpe/analysis/IXPE_Stats-Advice.pdf}} we accounted for the use of multiple phase bins and combined the phase-resolved polarization measurements through $\chi^{2} = \sum_{j=1}^{J} \left( {\Pi_{j}}/{\sigma_{j}} \right)^{2}$, with $2J = 16$ degrees of freedom (their Eq.~8), to evaluate the overall detection significance across the entire dataset.
This yields a global probability for the null hypothesis of $p_{\mathrm{glob}} \simeq 4 \times 10^{-4}$, corresponding to a significant detection at 3.3$\sigma$ (following the convention outlined in Sect.~3.4 of the recommendations).

\section{Pulsar geometry}
\label{sec:rvm}
 
To date, X-ray polarimetry with \ixpe has been reported for a broad range of accreting pulsars, including disk-fed systems such as \mbox{Her~X-1} \citep{2022NatAs...6.1433D} and \mbox{4U~1626$-$67} \citep{2022ApJ...940...70M}; wind-fed HMXBs with OB supergiant donors, including \mbox{Cen~X-3} \citep{Tsygankov22}, \mbox{Vela~X-1} \citep{Forsblom23}, \mbox{GX~301$-$2} \citep{2023A&A...678A.119S}, \mbox{SMC~X-1} \citep{Forsblom24}, \mbox{4U~1538$-$52} \citep{Loktev2025}, and \mbox{4U~1907$+$09} \citep{Zhou2025a}; as well as Be/X-ray binaries observed during outbursts, such as \mbox{EXO~2030+375} \citep{2023A&A...675A..29M}, \mbox{GRO~J1008$-$57} \citep{2023A&A...675A..48T}, \mbox{LS~V~+44~17} \citep{Doroshenko23}, \mbox{Swift~J0243.6$+$6124} \citep{Poutanen2024}, and \mbox{2S~1417$-$624} \citep{Zhou2025b}.

Typical luminosities during the published \ixpe observations are of the order $\gtrsim 10^{36}$~\lum in OB-supergiant HMXBs  and during Be outbursts, although lower values are occasionally reached in persistent low-$\dot{M}$ sources such as X~Persei \citep{2023MNRAS.524.2004M}. 
The observed polarization is modest, usually a few percent when averaged over the pulse, with phase-dependent enhancements and energy-dependent swings of the PA \citep{Poutanen2024Galax}. 
The detection of X-ray polarization from \src\ (see Sect.~\ref{sec:phres}) marks the first such measurement for an RSG-accreting XRP.

To constrain the geometry of \src, we fit the observed phase dependence of the PA using the rotating vector model (RVM; \citealt{Radhakrishnan1969,Meszaros1988,Poutanen2020}), as commonly applied in recent \ixpe studies of XRPs. 
In this model, the PA depends on the orientation of the magnetic and spin axes as
\begin{equation}
\tan(\mathrm{PA}-\chi_{\rm p}) = 
\frac{-\sin\theta\,\sin(\phi-\phi_{0})}
     {\sin i_{\rm p}\cos\theta - \cos i_{\rm p}\sin\theta\cos(\phi-\phi_{0})},
\end{equation}
where $\theta$ is the magnetic obliquity, $i_{\rm p}$ the inclination of the spin axis to the line of sight, $\chi_{\rm p}$ its position angle in the sky (assuming that the radiation is dominated by an ordinary mode) and $\phi_{0}$ is the reference phase. 

The application of the RVM implicitly assumes that the observed polarization is dominated by a single polarization mode (ordinary or extraordinary), such that the PA traces the projection of the magnetic axis on the sky. 
In the case of dominance of the extraordinary mode, the inferred position angle of the pulsar spin axis is $\chi_{\rm p}\pm90\degr$. In addition, the presence of an unpulsed phase-independent polarized component would introduce a constant offset in Stokes $(q,u)$ and could bias the inferred PA swing and geometry; this effect is argued to be important in XRPs LS~V~+44~17 / RX~J0440.9$+$4431 \citep{Doroshenko23} and Swift~J0243.6$+$6124 \citep{Poutanen2024}. 
For \src, the current data do not require such a component, but its presence cannot be excluded without additional observations.

We performed the RVM fitting using all phase bins.
For each bin, we evaluated the likelihood using the analytic probability density function of the measured PA derived by \citet{Naghizadeh-Khouei1993}:
\begin{equation}
G(\psi) =
\frac{1}{\sqrt{\pi}}
\left\{
\frac{1}{\sqrt{\pi}} + 
\eta \, {\rm e}^{\eta^{2}}\bigl[1+\mathrm{erf}(\eta)\bigr]
\right\}
{\rm e}^{-p_{0}^{2}/2},
\end{equation}
where $\eta = p_{0}\cos[2(\psi-\psi_{0})]/\sqrt{2}$, $\psi$ is the model PA, $\psi_{0}$ the observed PA, and $p_{0}=\mathrm{PD}/\sigma_{P}$ is the polarization signal-to-noise ratio. 
This approach allowed us to include even low-significance data points.

\begin{table}
\caption{Parameters of the RVM geometric solutions for binned and unbinned analyses.}
\centering
\begin{tabular}{lccc}
\hline
\hline
Parameter & \multicolumn{2}{c}{Binned} & Unbinned \\
 & Solution 1 & Solution 2 &  \\
\hline
$i_{\rm p}$ (deg)    & $101^{+11}_{-8}$ & $145^{+13}_{-14}$ & $144_{-15}^{+14}$ \\
$\theta$ (deg)       & $68_{-7}^{+5}$   & $47_{-17}^{+15}$  & $49_{-18}^{+16}$  \\
$\chi_{\rm p}$ (deg) & $16\pm8$   & $139^{+13}_{-12}$ & $138_{-12}^{+10}$ \\
$\phi_0$             & $0.47_{-0.02}^{+0.03}$ & $0.75\pm0.03$ & $0.76\pm0.02$ \\
\hline
\end{tabular}
\tablefoot{Position angle $\chi_{\rm p}=139\degr$ is equivalent to $-41\degr$ under the $180\degr$ modulo; other equivalent representations follow the same periodicity.}
\label{tab:rvm_solutions}
\end{table}

\begin{figure}
\centering
\includegraphics[width=\linewidth]{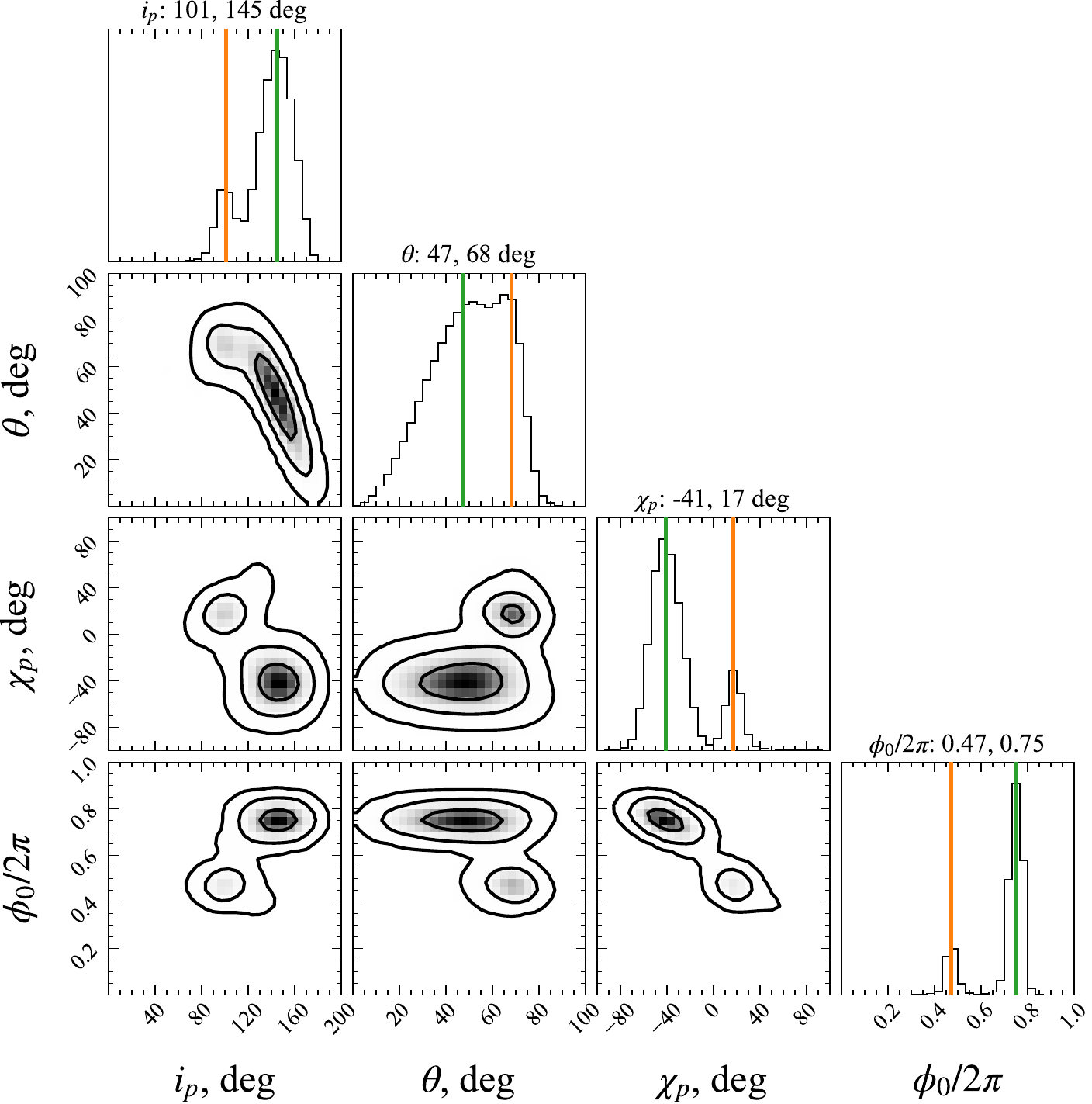}
\caption{Posterior distributions of the RVM parameters derived from the phase-resolved PAs (spectro-polarimetric analysis).
Contours correspond to 1$\sigma$, 2$\sigma$, and 3$\sigma$ CLs. The histograms show the normalized one-dimensional distributions for a given parameter derived from the posterior samples. Vertical lines in the diagonal panels mark the median values of the two posterior clusters.}
\label{fig:corner_rvm}
\end{figure}

We explored the parameter space with the Monte Carlo affine-invariant Markov chain (MCMC) sampler \textsc{emcee} \citep{Foreman-Mackey2013}, assuming an isotropic prior for the spin axis inclination $p(i_{\rm p})\propto\sin i_{\rm p}$ and uniform priors for the other parameters. 
The RVM fit yields two viable geometric solutions given in Table~\ref{tab:rvm_solutions} (see the posterior distributions of the parameters in Fig.~\ref{fig:corner_rvm} and the corresponding PA variations in Fig.~\ref{fig:phase_stack_rvm}e).  
To separate the distinct clusters of the posterior, we applied a two-component $k$-means clustering in the $(\chi_{\rm p},\phi_{0})$ parameter space, which isolates two clusters of the distribution. 
This behavior is expected given the limited polarization significance in several phase bins. 
In the corner plot, the two clusters exhibit peaks of different heights. 
Model comparison using the Akaike Information Criterion \citep[AIC;][]{Akaike1974}, computed as $\mathrm{AIC}=2k-2\ln L$ (with the same number of parameters $k$ for both solutions), yields $\Delta\mathrm{AIC}\approx3.5$, indicating only a marginal preference for Solution~2 and leaving the degeneracy unresolved.  

We performed complementary RVM analysis directly in the Stokes $(q,u)$ domain.
For each phase bin $i$, the model predicts
\begin{equation}
q_i^{\rm mod} = P_{0,i}\cos(2\psi_i), 
\qquad 
u_i^{\rm mod} = P_{0,i}\sin(2\psi_i),
\end{equation}
where $\psi_i$ is the RVM PA and $P_{0,i}$ is the PD, treated as an independent free parameter for each bin.
Assuming Gaussian uncertainties in the measured Stokes parameters
$(q_i^{\rm obs}, u_i^{\rm obs})$ with standard deviations
$\sigma_{i}$, the likelihood takes the form
\begin{equation}
\ln\mathcal{L} = -\frac{1}{2}\chi^2 + \mathrm{const},
\end{equation}
where
\begin{equation}
\chi^2  
= \sum_i 
\left[
\frac{(q_i^{\rm obs}-q_i^{\rm mod})^2}{\sigma_{i}^2}
+
\frac{(u_i^{\rm obs}-u_i^{\rm mod})^2}{\sigma_{i}^2}
\right].
\end{equation}

The posterior samples separate into two distinct geometric clusters in parameter space, which does not show a preferred configuration. The RVM provides an adequate description of the phase-resolved polarization data, with a reduced $\chi^2 \simeq 1.7$.

\begin{figure}
\centering
\includegraphics[width=\linewidth]{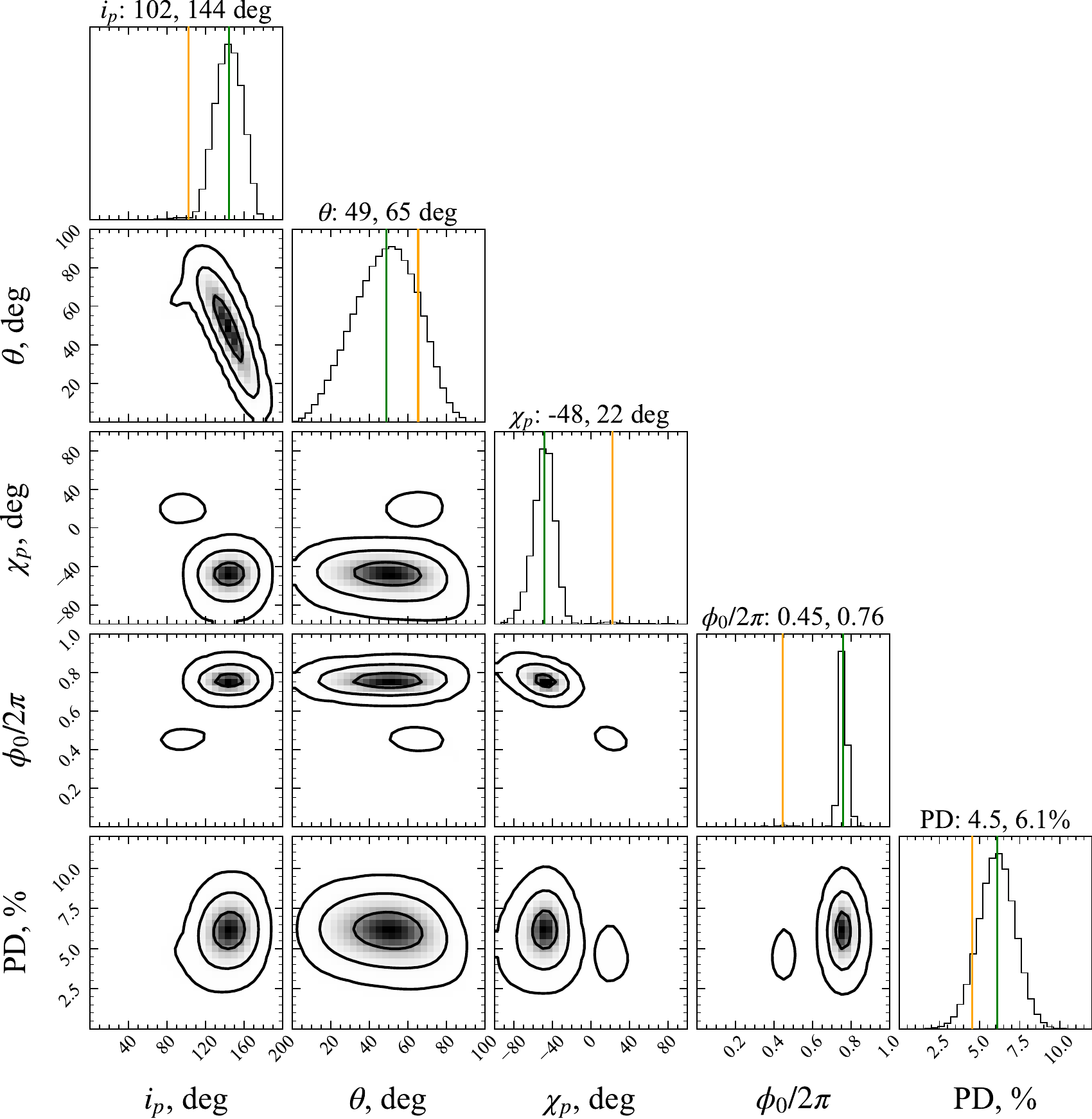}
\caption{Posterior distributions of the RVM parameters from the unbinned analysis.
Contours correspond to 1$\sigma$, 2$\sigma$, and 3$\sigma$ CLs. The histograms show the normalized one-dimensional distributions for a given parameter derived from the posterior samples.
Vertical lines in the diagonal panels mark the median values of the two posterior clusters.}
\label{fig:corner_rvm_unbinned}
\end{figure}

We performed an unbinned, event-by-event RVM fit (Fig.~\ref{fig:corner_rvm_unbinned}) using measured normalized Stokes parameters $(q_k,u_k)$ for each photon together with the per-event modulation factor $\mu(E)$ \citep{Gonzales2023}. 
In this case, the likelihood is written as a product of per-photon probabilities of the form
$f_k = (2\pi)^{-1}\bigl[1 + \mu_k P_0 \bigl(q^{\gamma}_k q^{\rm m}_k + u^{\gamma}_k u^{\rm m}_k\bigr)\bigr],$
where $(q^{\gamma}_k,u^{\gamma}_k)$ are the measured Stokes parameters of photon $k$, $(q^{\rm m}_k,u^{\rm m}_k)$ are the model predictions for its rotational phase, and $P_0$ is treated as a free parameter.

The \ixpe Level-2 event lists provide Stokes-like columns $Q_k$ and $U_k$, defined as $Q_k=2\cos(2\psi_k)$ and $U_k=2\sin(2\psi_k)$ (see Sect.~4.1 of the \ixpe\ Quick Start Guide\footnote{\url{https://heasarc.gsfc.nasa.gov/docs/ixpe/analysis/ixpe_quickstart.pdf}}), whereas the \citet{Gonzales2023} likelihood is formulated in terms of normalized quantities $q^{\gamma}_k=\cos(2\psi_k)$ and $u^{\gamma}_k=\sin(2\psi_k)$. We therefore used $q^{\gamma}_k=Q_k/2$ and $u^{\gamma}_k=U_k/2$ in the unbinned likelihood.

The unbinned posterior contains two previously considered solutions, but they are now  highly asymmetric in statistical weight. 
Information-criterion comparison shows that Solution~2 is strongly favored ($\Delta\mathrm{AIC}\approx 13$), allowing us to unambiguously identify it as the statistically preferred geometry. The geometric parameters of the preferred solution remain consistent with the binned results within their respective uncertainties (see Table~\ref{tab:rvm_solutions}), and the phase-independent polarization is characterized by ${\rm PD}=6.1\pm1.1$\%. This unbinned estimation of the PD explicitly accounts for the phase-dependent PA rotation, avoids cancellation of the Stokes vectors over the pulse, and yields a significant detection, unlike the phase-averaged analysis.

Thus, polarimetric analysis yields magnetic obliquity in the range $\theta \simeq 30\degr$--$65\degr$, corresponding to an intermediate viewing geometry. In this configuration, the angle between the line of sight and the magnetic axis varies significantly over the spin period, which does not favor simple or symmetric pulse profiles. 
The complexity of the flux profile is also driven by the properties of the emitting regions and the accretion process, including inhomogeneities such as clumpy stellar-wind accretion previously suggested for this source \citep{Bozzo2022}.

Such an intermediate obliquity is consistent with the range of values inferred for other accreting XRPs observed with \ixpe \citep[see, for a recent review,][]{Poutanen2024Galax}, which span from nearly aligned to nearly orthogonal rotators. 
At the same time, \src\ complements the existing \ixpe\ sample by providing the first geometric constraints for an extremely slowly rotating XRP accreting from the wind of a red supergiant companion.

\section{Conclusions}
\label{sec:conclusions}

In this work, we present the first X-ray polarimetric study of the red supergiant XRP \src\ with \ixpe.  
Our main results can be summarized as follows.

\begin{enumerate}
    \item The \ixpe observation confirms the presence of very slow coherent pulsations with a spin period of $P_{\rm spin}=5.49\pm0.05$~h.  
    The 2--8~keV pulse profile has two maxima with multiple subpeaks and exhibits a PF of $55\pm1$\%.

    \item No significant phase-averaged polarization is detected in the 2--8~keV band, with an MDP$_{99}$ of 4.9\%.

    \item The phase-resolved analysis identifies one phase bin in which the PD exceeds its MDP$_{99}$.  
    At pulse maximum, the measurement yields ${\rm PD}=10.2_{-3.0}^{+3.1}$\% and ${\rm PA}=83\degr\pm9\degr$.
    Together with lower-significance measurements in the remaining bins, these results trace a smooth rotation of the PA across the pulse by $\approx 150\degr$.  
    When all phase bins are combined in a joint statistical test, the overall detection significance reaches $3.3\sigma$.

    \item Fitting the phase dependence of the binned polarimetric data with the RVM yields two plausible geometric solutions. The unbinned analysis indicates that one of them statistically dominates, and we therefore adopt it as the preferred geometry, yielding a phase-independent PD of $6.1\pm1.1$\% in the 2--8~keV band.
\end{enumerate}

\begin{acknowledgements}
This work reports observations obtained with the Imaging X-ray Polarimetry Explorer (IXPE), a joint US (NASA) and Italian (ASI) mission, led by Marshall Space Flight Center (MSFC). The research uses data products provided by the IXPE Science Operations Center (MSFC), using algorithms developed by the IXPE Collaboration (MSFC, Istituto Nazionale di Astrofisica - INAF, Istituto Nazionale di Fisica Nucleare - INFN, ASI Space Science Data Center - SSDC), and distributed by the High-Energy Astrophysics Science Archive Research Center (HEASARC). This research also has made use of the \nustar Data Analysis Software (NUSTARDAS) jointly developed by the ASI Science Data Centre (ASDC, Italy) and Caltech. We are grateful to the \ixpe\ and \nustar\ teams for the approval and rapid scheduling of the observations.
AS acknowledges support from the Jenny and Antti Wihuri Foundation (grant no. 00240331).
This research was supported by the International Space Science Institute (ISSI) in Bern, through International Team project 25-657 `Polarimetric Insights into Extreme Magnetism' and the  Research Council of Finland Centre of Excellence in Neutron-Star Physics (grant 374064). We thank the anonymous referee for their careful reading of the paper and thoughtful suggestions, which helped improve the clarity of the results.

\end{acknowledgements}

\bibliography{allbib}
\bibliographystyle{aa}
\end{document}